\documentclass[twocolumn,showpacs,english,prl,preprintnumbers,amsmath,amssymb,floatfix]{revtex4}

\usepackage{subfiles}

\usepackage[T1]{fontenc}
\usepackage[latin9]{inputenc}
\usepackage{dcolumn}
\usepackage{bm}
\usepackage{graphicx}
\usepackage{color}
\usepackage{esint}
\usepackage{babel}
\usepackage{amsmath,amssymb,amsthm,mathrsfs,amsfonts,dsfont}
\usepackage{slashed}
\usepackage{enumerate}
\usepackage{simplewick}

\begin{document}

\title{
Non-ergodic extended states in the SYK model
}

\author{ 
T.~Micklitz, Felipe Monteiro
}
\affiliation{
Centro Brasileiro de Pesquisas F\'isicas, Rua Xavier Sigaud 150, 22290-180, Rio de Janeiro, Brazil 
} 

\author{Alexander Altland}
\affiliation{Institut f\"ur Theoretische Physik, Universit\"at zu K\"oln, Z\"ulpicher Str. 77, 50937 Cologne, Germany
}

\date{\today}

\pacs{05.45.Mt, 72.15.Rn, 71.30.+h}

\begin{abstract}
We analytically study spectral correlations and many body wave functions of an
SYK-model deformed by a random Hamiltonian diagonal in Fock space. Our main result
is the identification of a wide range of intermediate coupling strengths where the
spectral statistics is of Wigner-Dyson type, while wave functions are non-uniformly
distributed over Fock space. The structure of the theory
suggests that such manifestations of non-ergodic extendedness may be a prevalent
phenomenon in many body chaotic quantum systems.
\end{abstract}

\maketitle

{\it Introduction:---}In recent years, classifications of many body quantum systems
as either `ergodic', or `many body localized' (MBL) have become mainstream. This
reflects the discovery of a growing number of systems supporting MBL
phases~\cite{BaskoAleinerAltshuler,Mirlin,fermions1,bosons1,bosons2,bosons3,spin1,spin4,spin6,spin7,Imbrie,GornyiMirlinPolyakov17}
and naturally extends the distinction between single particle ergodic and Anderson
localized systems to many body quantum disorder. However, recently, we are seeing
mounting evidence~\cite{NonE_Ex,RPmodel1,Kravtsov1,Kravtsov2,Mirlin1,TorresSantos17,faoro,Biroli12,Biroli18} that the
above dichotomy may be too coarse to capture the complexity of chaotic many body
systems. Specifically, recent work has put the focus on the study of statistical
properties of many body wave functions. It has been reasoned that, sandwiched
between the extremes `ergodic' and `many body localized', there might exist
intermediate phases of \emph{non-ergodic extended (NEE) states}, i.e. quantum states
different from localized in that they have unbounded support, and different from
ergodic in that their amplitudes are not uniformly distributed. One reason why this
option comes into focus only now is that standard tools in diagnosing chaos ---
spectral statistics applied to systems of  small size of $\mathcal{O}(10^1)$ physical
sites --- are too coarse to resolve the spatial structure of quantum states in
Fock space. Indeed, the above indications are indirect in that they are based on
numerical and analytic work on disordered graphs with high coordination numbers,
artificial systems believed to share key characteristics with genuine random Fock
spaces. The complexity of the matter shows in that, even for this synthetic system,
there is a controversy between work suggesting an NEE phase~\cite{NonE_Ex,RPmodel1,Kravtsov1,Kravtsov2} and other refuting
it~\cite{Mirlin1}.

\begin{figure}
\centering
\includegraphics[width=8.5cm]{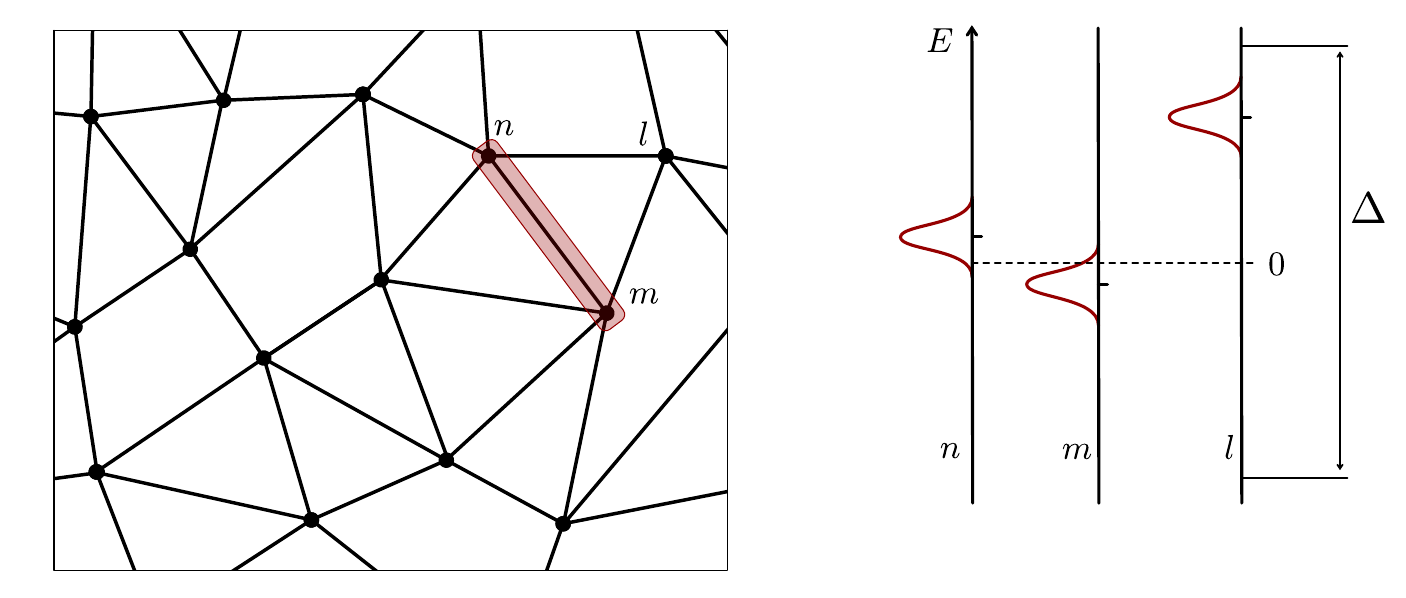}
\caption{\label{fig4} 
 Left: Cartoon of Fock space sites $n,m,l,\dots$ (indicated by dots) connected by 
hopping operator ${\cal P}$ (solid lines).  For
$\Delta\gg 1$ exceeding the band width of the unperturbed model, 
one may approach the problem  perturbatively,
i.e. taking the isolated eigenstates of levels $v_n,v_m,v_l$ as a starting point.
 The hybridization leads to level broadening, $\kappa$, of resonant neighbors 
 (indicated by hatched link) which 
 have both energies $v_n,v_m\lesssim 1$ within the SYK band. 
Right side: Typical energy distributions of Fock-space neighbors connected by ${\cal P}$.
The hybridization does in general not 
generate overlap between neighboring sites. For $\Delta<N^2$ wave functions
are thus extended ($\to$ Wigner-Dyson statistics) yet confined to only a
fraction $\sim 1/\Delta^2$ of the total Fock space.}

\end{figure}

In this paper, we present a first principle analytic description of NEE states in a deformed version of the SYK model~\cite{SYK1,SYK2}. The standard
SYK model is a system of $2N\gg 1$  Majorana fermions,
$[\chi_i,\chi_j]_+=2\delta_{ij}$, governed by the interaction Hamiltonian
\begin{align}
\label{syk}
\hat H_0
&=
{1\over 4!}\sum_{i,j,k,l=1}^{2N}
J_{ijkl} \hat\chi_i \hat\chi_j \hat\chi_k \hat\chi_l , 
\end{align}
where the coupling constants are drawn from a Gaussian distribution, $\langle
|J_{ijkl}|^2 \rangle= 6J^2/(2N)^3$,  and the constant $J$ defines the effective band width of the system as $\gamma=\frac{J}{2}(2N)^{1/2}$~\cite{GG2}. 
The model~\eqref{syk} is known to be in an ergodic phase with eigenfunctions
uniformly distributed in Fock space~\cite{GG2,Cotler}. To make the situation more interesting, we generalize the Hamiltonian to $\hat H = \hat H_0 + \hat H_V$, where
\begin{align}
     \label{syk+V}
     \hat H_V &= 
     \gamma\sum_{n}^D v_n |n\rangle \langle n|, 
 \end{align} 
 is a sum over projectors onto the occupation number eigenstates $|n\rangle =
|n_1,n_2,\dots,n_N\rangle$,  $n_i=0,1$, of a system of complex fermions  $c_i =
\tfrac{1}{2}(\chi_{2i-1}+ i \chi_{2i})$, $i=1,\dots, N$ defined via the Majorana operators. The
coefficients $v_n$ can be chosen to represent any operator diagonal in the occupation
number basis, $\{|n\rangle\}$,  pertaining to  a fixed one-body basis. For example,
any one-body operator~\cite{SYK_GG,Shepelyansky17}, $\hat H_0=\frac{1}{2}\sum_{i,j} J_{ij}\hat \chi_i \hat
\chi_j$, can be diagonalized in the fermion representation and described in this way.
However, for our discussion below it will be sufficient to consider realizations of
maximal entropy with coefficients $v_n$ drawn from a box distribution of width
$\Delta$ symmetric around zero. In this way $\Delta$ sets the effective strength of
the coupling in units of the SYK bandwidth, and in the limit of asymptotically
large $\Delta$ enforces Fock space localization in states $n$ with energies $v_n$.
The Hamiltonian $\hat H_0$ perturbs this `Poisson limit' via transitions
$|n\rangle\rightarrow|m\rangle$ between states nearby in Fock space. (The two-body
$\hat H_0$ changes the occupation of a state $|n\rangle$
by at most four, and it preserves the number parity, where we focus
on even parity states throughout.) It does so via only an
algebraically small number $\sim N^4\sim
\ln(D)$ of independent matrix elements, and thus defines an operator with strong
statistical correlation. However, we will see that $\hat H_0$ is very efficient in
introducing many body chaos, as evidenced by the onset of Wigner-Dyson (WD) spectral
statistics, including for values $\Delta\gg 1$ where the diagonal still dominates.
Our main objective is  to explore the profile of the many body wave functions in this
setting.

\emph{Qualitative picture:---}Before turning to the quantitative analysis of the
problem, let us outline an intuitive picture of non-ergodic wave function
statistics. Let us work in dimensionless units,
where the SYK bandwidth $1\sim JN^{1/2}$ is set to unity, or $J\sim N^{-1/2}$.
Consider a situation where the strength of the diagonals, $\Delta \sim N^\alpha$,
$\alpha>0$ parametrically exceeds the band width. In this case, we have a situation
where the `hopping' in Fock space  induced by the  SYK Hamiltonian does not effectively hybridize the majority of the $\sim N^4$ states, $m,l,\dots$,
neighboring a given $n$, cf. Fig.~\ref{fig4}. With the
characteristic hopping amplitude $t\sim J N^{-3/2}\sim N^{-2}$, a self consistent
golden rule argument may be applied to estimate the residual smearing, $\kappa$, of
$n$ as $\kappa \sim |t|^2 \left(N^4
\frac{\kappa}{\Delta}\right)\frac{1}{\kappa}\sim \frac{1} {\Delta}\sim N^{-\alpha}$,
where the term in parentheses is the number of neighbors that are in resonance, and
$\sim \kappa^{-1}$ is the broadened energy denominator. The effective hybridization of two nearest neighbors requires overlap of
their smeared levels, a condition satisfied only by a fraction $ 
\frac{\kappa}
{\Delta}\sim \Delta^{-2}$ of  neighbors. From this argument we infer
that typical wave functions occupy only a number $D/\Delta^2\sim D/N^{2\alpha}$
of the available $D$ sites in Fock space. We also note that for $N^4 /\Delta^2=N^{4-2\alpha}\sim 1$
the number of resonant neighboring levels becomes of $\mathcal{O}(1)$. This is when we
expect the
wave functions to fragment and a transition to the Poisson regime to take place.


\begin{figure}
\centering
\includegraphics[width=5.5cm]{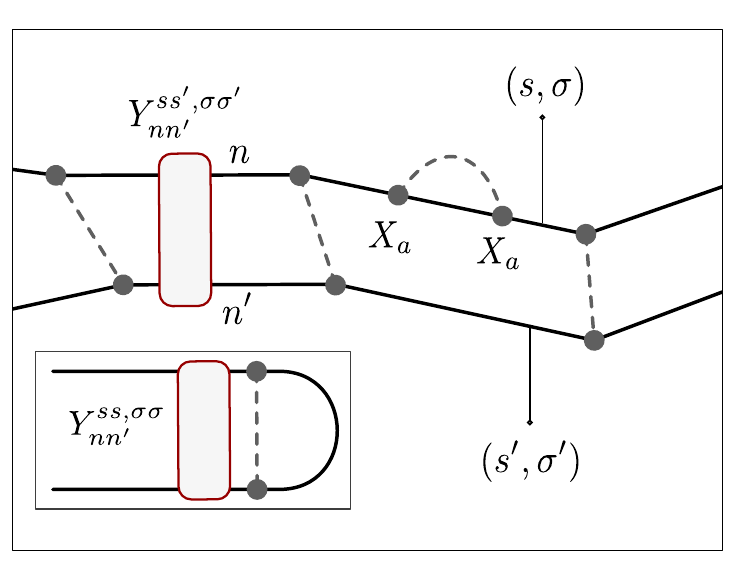}
\caption{\label{fig3} 
The scattering of wave function amplitudes in Fock space. Variables
 $Y^{ss',\sigma\sigma'}_{nn'}$ describe the correlated propagation of resolvents
 (solid lines) labeled by a conserved index $(s,\sigma)$. Scattering processes
 (indicated by dots) can be distinguished into those dressing propagators by `self
 energies' (dashed lines connecting same resolvent) and vertex contributions (dashed
 lines connecting different resolvents). Hatched regions summarize repeated,
 ladder-diagrams of vertex contributions and define the slow modes in the system.
 Inset: Self consistency equation for self-energy Eq.~\eqref{eq:MeanFieldEquation}. }
\end{figure}
{\it Matrix integral representation:---}To obtain a more quantitative picture, we
start from a first quantized representation, where the  Hamiltonian $\hat H$ is
considered as a sparse matrix acting in a huge Fock-space. This perspective is
complementary to that of the more conventional many body
$G\Sigma$-formalism~\cite{SYK1} probing the physics of collective fluctuations close
to the ground state. Formulated in this language, the problem becomes one of random
matrix diagonalization and methods such as the powerful supersymmetry technique, originally designed to solve single particle
hopping problems, become applicable. Specifically, the occupation number basis
$\{|n\rangle\} $ plays a role analogous to the position basis of a fictitious quantum
state and $\hat H_0$ and $\hat H_V$ act as `hopping' and  `on-site potential'
Hamiltonians, respectively.  Within the first quantized approach, information
on the statistics of the many body wave functions $|\psi\rangle$ at the band center,
$\epsilon_\psi=0$ (generalization to generic energies is straightforward but omitted
for simplicity), is contained in the matrix elements of the resolvent, $
G^{\pm}_{nn'}=\big\langle n |\pm i\delta - \hat H)^{-1}|n'\big\rangle $.
Specifically, the $q$th moment is defined as $I_q \equiv
\frac{1}{\nu_0}\sum_{n}
\left\langle| \langle \psi|n\rangle |^{2q}\delta(\epsilon_\psi)\right\rangle$,
where $\langle \dots\rangle$ denotes averaging over the randomness in the model, and
$\nu_0 = \langle \sum_\psi \delta(\epsilon_\psi)\rangle$ is the density of states
in the band center. 
 Using the eigenfunction
decomposition $G^+_{nn}= \sum_\psi |\langle \psi|n\rangle |^2 (i\delta -
\epsilon_\psi)^{-1}$, this can be expressed as $I_q=-
\frac{1}{\pi \nu_0}\lim_{\delta\to 0}(2i\delta)^{q-1} \sum_n
\mathrm{Im}\,G^+_{nn}G^{+(q-1)}_{nn}$~\cite{Wegner79}, where the last equality relies on  the absence of degeneracies $E_\psi\not=E_{\psi'}$,
for $\psi\not={\psi'}$ in a disordered system. (For completeness, we  apply the same
setup to compute the eigen\emph{value} statistics and diagnose Wigner--Dyson or
Poisson statistics. See supplementary material for details.) Our principal workhorse
in computing the realization average of these expressions is   an exact integral
representation $\langle I_q\rangle=\partial_\beta \partial^{q-1}_\alpha\int dY
e^{-S(Y,\alpha,\beta)}$. Here, the integration variables
$Y=\{Y_{nn'}^{ss',\sigma\sigma'}\}$ are $2\times2\times D$-dimensional matrices which
on top of the Fock space index $n$ contain an index $s,s'=\pm$ labeling  advanced and
retarded states, and a two-component index $\sigma,\sigma'=\mathrm{b,f}$
distinguishing between commuting ($Y^{\mathrm{bb}},Y^{\mathrm{ff}}$) and Grassmann
valued ($Y^{\mathrm{bf}},Y^{\mathrm{fb}}$) matrix blocks~\cite{fn1}. This
`supermatrix structure'~\cite{SuppMat} is required to cancel unwanted fermion
determinants appearing in the computation of purely commuting or anti-commuting
matrix integrals. (We cannot use replicas to achieve determinant cancellation because
the analysis will involve one non-perturbative integration, not defined in the
replica formalism.)

Referring for a derivation of the above integral, and the discussion of the source
parameters $\alpha,\beta$ required to generate the wave function moments to the
supplementary material, the action $S(Y)\equiv S(Y,0,0)$ of the field integral is
given by
\begin{align}
 \label{sm}
S(Y)
&=
-{1\over 2}
{\rm STr}( Y \mathcal{P}^{-1} Y )
+
{\rm STr}\ln\left(i\delta  
\sigma_3 
-\hat H_V 
+
i\gamma Y
\right),
\end{align}
where  $\mathrm{STr}(X)\equiv \sum_{n,s,\sigma}(-)^\sigma X_{nn}^{ss,\sigma\sigma}$
is the canonical trace operation for supermatrices~\cite{Efetov}. To understand the structure of
the action, notice that the  Green functions describe the propagation of wave
functions subject to random scattering in Fock space. Contributions surviving the
configuration average are correlated as indicated in Fig.~\ref{fig3}. The first term
in the action describes how the pair amplitudes $Y_{n,n'}^{ss',\sigma\sigma'}$
represent the propagation of two such states, specified by a doublet of indices
$(n,s,\sigma)$ and $(n',s',\sigma')$. It is defined by an operator $\mathcal{P}$,
which acts as $\mathcal{P}Y\equiv \frac{1}{{\cal N}}\sum_a X_a Y X^\dagger_a$ 
where $\mathcal{N}=
\binom{2N}{4}$, i.e.
the multiplication of the two states represented by $Y$ by the  Majorana product
operators contained in the Hamiltonian, where $X_a\equiv
\chi_i\chi_j\chi_k\chi_l$, and the shorthand $a=(i,j,k,l)$ is used. The second term
couples the $Y$-matrices to the fermion propagator effectively describing the
propagation in-between SYK-scattering events, where $(\sigma_3)^{ss'}=
(-)^s\delta^{ss'}$ does the bookkeeping on causality. 

\emph{Stationary phase approach:---}Our strategy is to evaluate the matrix
integral by  stationary phase methods backed by excitation gaps present in the limit $\delta\to 0$. The structure of the action suggests to
look for solutions of the stationary phase equations $\delta_{\bar Y}S(\bar Y)=0$
diagonal in Fock space $ Y_{nn'}= Y_n \delta_{nn'}$. Physically, this
restriction means that for a fixed realization of the diagonals $v_n\not=0$ phase
coherence of the pair propagation requires $n=n'$ in the representation of
Fig.~\ref{fig3}. The stationary phase equation then assumes the form
\begin{align}
    \label{eq:MeanFieldEquation}
     Y_n = i\sum_m \Pi_{nm}   \frac{1}{ i\frac{\delta}{ \gamma}\sigma_3-  v_m +   iY_m},
\end{align}
where the projection of the pair-scattering operator $P_\mathrm{d}
\mathcal{P}P_\mathrm{d}\equiv \Pi$ on the space of diagonal matrix configurations
acts on  diagonal configurations as $(\Pi X)_n=
\sum_m \Pi_{nm}X_m=\frac{1}{{\cal N}}\sum_{a,m} |(\hat X_a)_{mn}|^2 X_m$. The solution of
the equation now essentially depends on the structure of this operator. We first note
that the operators $\hat X_a$ change at most four  of the $N$ binary
occupation numbers contained in $n$, implying that $\Pi$ is a local
hopping operator in the space of $n$-states. The permutation symmetry inherent to the
sum over all configurations $a=(i,j,k,l)$ further implies that the hopping strengths,
$\Pi_{nm}=\Pi_{|n-m|}$ depend only on the occupation number difference between Fock
space states, where a straightforward counting procedure yields 
$\Pi_0=N(N-1)/2\mathcal{N}$, $\Pi_2=4(N-2)/\mathcal{N}$, and
$\Pi_4=16/\mathcal{N}$, and all other matrix elements vanish. Armored with this result, we interpret the r.h.s. of the mean
field equation Eq.~\eqref{eq:MeanFieldEquation} as a sum over a large number  of
terms, which are effectively random due to the presence of the coefficients $v_m$. In
this way, $ Y_n(v)$ becomes a random variable depending on the realizations
$v=\{v_m\}$.

The structure of the mean field equation, and the transition rates $\Pi_{nm}$
identifies the  components $Y^{ss}_n$ as the self energies dressing the retarded
($s=+$) and advanced ($s=-$) Fock space propagators (also cf. inset of
Fig.~\ref{fig3}.). The solutions $ Y_n$ are obtained as sums over  large numbers of
random contributions which for small $\Delta$ implies a self averaging property, $Y_n
\simeq \langle Y_n\rangle_v\equiv Y_0$, where the r.h.s. denotes the average over the
independent distribution over $v_m$. Ignoring the imaginary part of $Y_n$ (which does
no more than inducing a weak shift $v_n \to v_n + \mathrm{Im} Y_n\simeq v_n$ of the
random energies), and averaging  $v$ over a box distribution, $\langle
\dots\rangle_v=\prod_m\int_{-\Delta/2}^{\Delta/2} {dv_m\over \Delta}(\dots)$, we
obtain $Y_0=\kappa \sigma_3$, where the self energy, $\kappa$,
obeys the equation $\kappa =(2/\Delta) \arctan(\Delta/2\kappa)$. The solution smoothly
interpolates between $\kappa\simeq 1$ for the weakly perturbed model, $\Delta \ll 1$
and $\kappa \simeq \pi/\Delta$ for $\Delta \gg 1$. In accordance with the qualitative 
discussion above, this decay reflects that for $\Delta\gg 1$ the majority of sites
neighboring a fixed $n$ are off-resonant and decouple from the self energy. 
We also note that the averaged density of states  
$\nu_0=-\mathrm{Im}\,\langle \mathrm{tr}(G^+)\rangle= D\kappa/\pi \gamma$  shows the
same behavior. Before proceeding, let us ask when the above approximations  breaks down  and the
stationary solutions become strongly fluctuating in the sense $
\mathrm{var}(Y_n)>Y_0^2$. Assuming that $Y_m\simeq Y_0$ on the
r.h.s. of Eq.~\eqref{eq:MeanFieldEquation},  a straightforward calculation leads
to  $\mathrm{var}(Y_n)\simeq
\frac{10\pi}{N^4\kappa^2}\mathcal{F}(\Delta/2\kappa)$, where $\mathcal{F}(x)$ is a
function monotonically increasing from $\mathcal{F}(0)=0$ to
$\mathcal{F}(x)=\mathcal{O}(1)$ at $x\sim 1$ before decaying as $\mathcal{F}(x)\sim
1/x$ at $x\gg 1$~\cite{fn2}. A balance $
\mathrm{var}(Y_n)\sim Y_0^2$ is reached  when
$\kappa^2 \sim \Delta^{-2}\sim \frac{1}{N^4 \kappa^2}\frac{\kappa}{\Delta}\sim
N^{-4}$, where $\kappa \sim \Delta^{-1}$ was used. This shows that only for disorder
strength $\Delta > \Delta_{\mathrm{P}} \sim N^2$ parametrically  larger than the
bandwidth, the homogeneity of the stationary phase configuration in Fock space gets compromised.
This observation  is one of the most important
results of this Letter. As we will demonstrate in the following, it provides the
basis for the analytical extraction of wave functions and spectra.

\emph{Wave function statistics:---}In the limit $\delta\to 0$, $Y_0=\kappa
\sigma_3$ is but one element of a manifold of stationary solutions, $Y_0 = \kappa T
\sigma_3 T^{-1}\equiv \kappa Q$, where $T=\{T^{ss',\sigma\sigma'}\}$ is a $4\times 4$
rotation matrix in advanced-retarded and super-space. The absence of Fock-space
indices implies $[\mathcal{P},Q]=0$, which in combination with $Q^2=\openone$ means
that the first term in Eq.~\eqref{sm} is independent of $T$. We conclude that the
stationary phase action of the matrix integral is given by
\begin{align}
  \label{sm_action}
S[Q]
&=
{\rm STr\ln}\left(
i\delta\sigma_3 -\hat H_V + i\gamma \kappa\, Q
\right).
\end{align}
This  action is known to describe~\cite{Ossipov} the Rosenzweig-Porter (RP)
model~\cite{RP}: a $D$-dimensional Gaussian random matrix ensemble perturbed by a
fixed diagonal, $\hat H_V$. We thus conclude that for diagonals with  $\Delta<
\Delta_{\mathrm{P}}$ the deformed SYK model and this much simpler model are in the
same universality class.   The first step 
 of the computation of the wave function statistics~\cite{Ossipov} based on~\eqref{sm_action}  
 is the integration over the matrix $T$. This integration is
not innocent, because the  $2\times 2$ block $T^\mathrm{bb}$ defines a non-compact
integration manifold~\cite{Efetov}. The convergence of the corresponding integral is safeguarded
only by the infinitesimal symmetry breaking parameter $i\delta$, and integration over
$T$~\cite{SuppMat} indeed produces a singular factor 
$\delta^{-q+1}$ canceling
the $\delta$-dependence  in the definition of the wave
function moments, and leading to the result
\begin{align}
     I_q=\frac{q!}{\nu_0^q}\sum_n\left\langle
     \nu(n)^q\right\rangle_v,\quad \nu(n)\equiv \frac{\nu_0}{D(v_n^2+\kappa^2)}.
 \end{align} 
Intuitively, the r.h.s. contains the $q$th moments of local Green function matrix
elements, with energy denominators broadened by the self energy $\kappa$. It is
straightforward to average this expression over the box distribution of the
individual $v_n$ and obtain
\begin{align}
\label{i_q}
I_q
&=
-(-2)^q q D^{1-q} \partial_{y_0^2}^{q-1} 
(1/y_0 \Delta) 
\arctan\left(
\Delta/2y_0
\right).
\end{align}
For $\Delta\ll 1$ smaller than the SYK bandwidth,  this asymptotes to the random
 matrix result $I(q)=q! (D/2)^{1-q}$, demonstrating a uniform state distribution. In
 the opposite case,  $\Delta\gg 1$, $y_0=\pi/\Delta$ and the moments
$
I_q
=
(2\pi^2)^{1-q}
 q(2q-3)!! 
  \Delta^{2(q-1)}
 D^{1-q}, 
$ 
show power law scaling in $\Delta$. Finally, for  $\Delta \sim N^\alpha$ the wave
 functions become non-ergodic $I_q \sim [D/ N^{2\alpha} ]^{1-q}$, and now only occupy
 a $\sim 1/N^{2\alpha}$ fraction of Hilbert space, in line with the qualitative
 discussion above. 
In Fig.~\ref{fig2}, these predictions are compared to wave function moments obtained by  exact diagonalization for $N=13$ as a function of the deformation parameter (main panel), or as a function of system size $N=7,\dots,13$ at fixed deformation (lower left panel). The figure demonstrates excellent, and parameter free agreement with the analytic result. 

 The figure also confirms the statement that throughout the entire window $\Delta < \Delta_P$, or $0\le \alpha  <2$,  the spectral
 statistics remains Wigner-Dyson like. This is probed  by comparing the 
relative, or Kullback-Leibler entropies~\cite{KullbackLeibler}
 ${\rm KL}(p|q)\equiv\sum_k p_k \ln(p_k/q_k)$ between  
the numerically obtained moments $q_k$ and the Wigner-Dyson, or Poisson distribution
 $p_k$, respectively. 
 The upper inset of Fig.~\ref{fig2} shows 
  that the change between the two statistics takes place at the deformation strength analytically predicted as $\Delta\sim \Delta^P\simeq 120$, beyond which both saturation of the wave function moments\cite{fn3}, and the level statistics indicate Poissonian behavior.   

  Conceptually, the robustness of spectral correlations follows from the equivalence
 (SYK $\stackrel{\Delta < \Delta_P}\sim $ RP), the latter being a model demonstrating
 the strong resilience of a single random matrix against perturbations on its
 diagonal. The domain of the above equivalence is limited by both the deformation
 strength of SYK $\Delta\lesssim N^2$, and the width of the probed  spectrum
 $\epsilon\lesssim \delta N^2$, where $\delta$ is the many body level spacing
 \cite{ErgodSYK}. Outside this window, for $\Delta\gtrsim N^2$, the theory predicts a
 fragmentation of the Fock space homogeneous mean field (equivalent to the
 fluctuations of a single random matrix ensemble) into inhomogeneous stationary
 configurations, $\kappa\to \kappa_n$. On the background of this inhomogeneous
 configuration one may construct a  lattice field theory which indeed predicts a Fock
 space localization transition at $\Delta\sim
 \Delta_\mathrm{P}$~\cite{AMTbp}. Finally, models of the perturbation different from the identically distributed $v_n$, lead to similar results. Specifically, a random one-body term, $\hat
 H_1\equiv \sum_{j=1}^N \eta_{2j-1}\eta_{2j}v_j$ is equivalent to $\hat H_V$ with
 statistically correlated $v_n(\{v_j\}\})$. Referring to Ref.~\cite{AMTbp} for
 details, this leads to similar scaling over a slightly higher tolerance window,
 $\Delta_\mathrm{P}\lesssim N^{9/4}$.

\begin{figure}
\centering
\includegraphics[width=8.4cm]{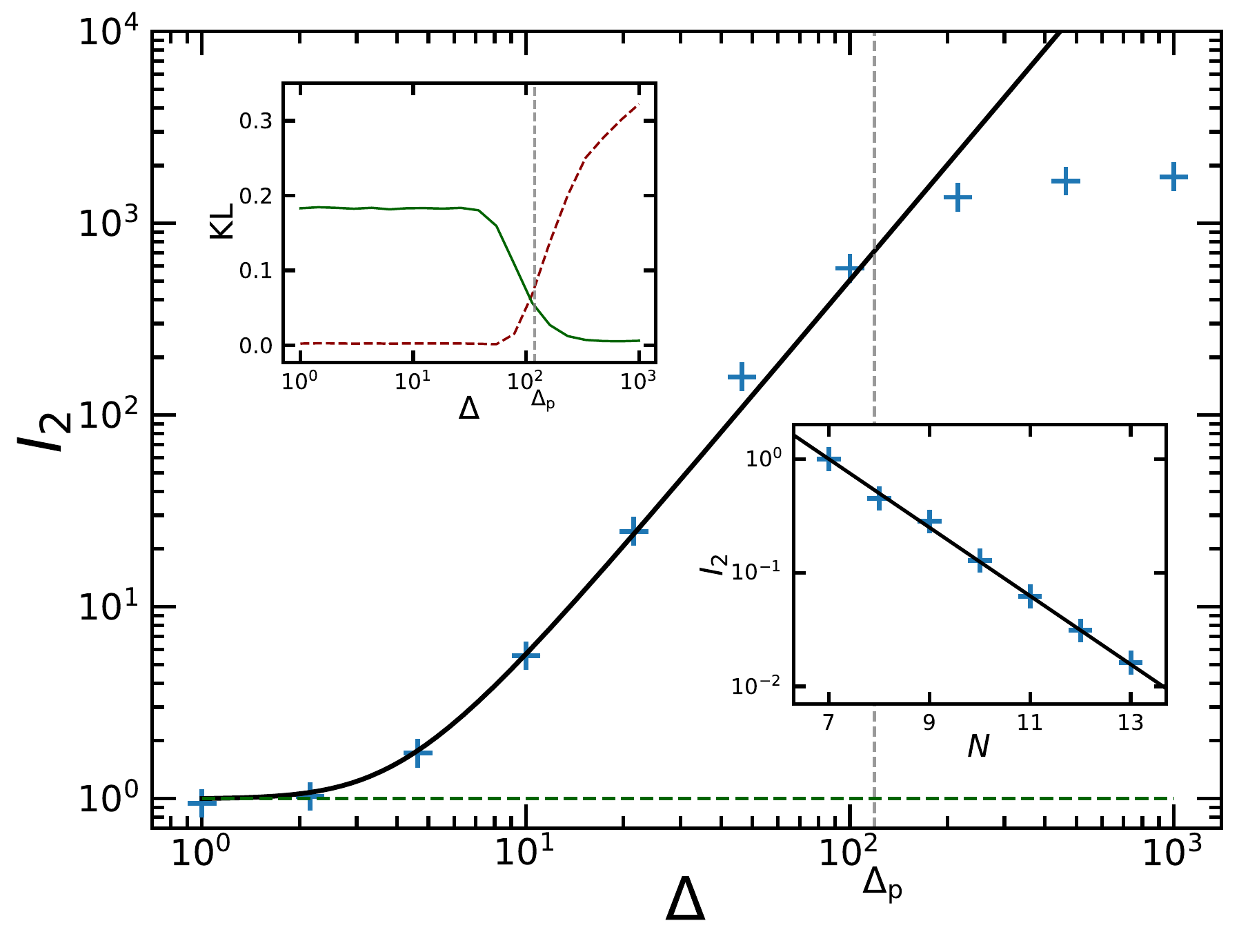}
\caption{\label{fig2} 
Inverse participation ratio as a function of $\Delta$, normalized by $I_2(\Delta=0)$, from exact diagonalization $N=13$; the 
analytical prediction Eq.~\eqref{i_q} is indicated by the solid line. Left inset: relative entropy (Kullback-Leibler) between 
numerical and Wigner Dyson (dashed), resp., Poisson (solid) distributions. Right inset: 
Inverse participation ratio as a function of $N$, normalized by $I_2(N=7)$, from exact diagonalization at $\Delta=10$;  
solid line is the analytical prediction from Eq.~\eqref{i_q}~\cite{fn3}.}
\end{figure}

{\it Summary and discussion:---}The  model considered in this paper defines the
perhaps simplest many body system showing a competition between Fock space
localization and ergodicity. We are seeing unambiguous evidence that the passage
between the two limits is not governed by a single many body localization transition
but contains a parametrically extended intermediate phase characterized by a coexistence of
Wigner-Dyson spectral statistics and non-trivial extension of wave functions over
Fock space.
 Methodologically, this phenomenon emerged as the result of a competition:
the `hopping' in Fock space generated by the SYK two-body interaction stabilized a
uniform mean field against the `localizing' tendency of the Fock-space diagonal
operator, $\hat H_v$.  We have identified an intermediate regime, where the corresponding low energy theory is governed by a homogeneous fluctuation mode, $T_0$, acting on top of a background containing inhomogeneous energy denominators. This mechanism appears to be of rather general nature and
makes one suspect that non-ergodic wave function statistics in coexistence with RMT
spectral correlations could be a  more frequent phenomenon than previously
thought.

{\it Acknowledgements:---} Discussions with D. Bagrets, A. Kamenev, and H. Wang are gratefully acknowledged. T.~M. and F. M.~acknowledge financial support 
by Brazilian agencies CNPq and FAPERJ. Work funded by the Deutsche Forschungsgemeinschaft (DFG, German Research Foundation) - 
Projektnummer 277101999 - TRR 183 (project A03).
%


\clearpage

\title{
Non-ergodic extended states in the SYK model: supplementary material
}

\author{ 
 T.~Micklitz, Felipe Monteiro
}
\affiliation{
Centro Brasileiro de Pesquisas F\'isicas, Rua Xavier Sigaud 150, 22290-180, Rio de Janeiro, Brazil 
} 

\author{Alexander Altland}
\affiliation{Institut f\"ur Theoretische Physik, Universit\"at zu K\"oln, Z\"ulpicher Str. 77, 50937 Cologne, Germany
}

\pacs{05.45.Mt, 72.15.Rn, 71.30.+h}

\begin{abstract}

In this supplemental material we give details on the derivation of the matrix integral representation  
and the calculation of moments of the wave function and level statistics.

\end{abstract}

\maketitle

\section{SUSY Matrix integral in a nutshell}

In this section we provide a concise yet self contained derivation of the supersymmetric matrix integral representation for the computation of SYK Green functions. 

{\it Generating function:---} The general starting point for such constructions is
the Gaussian integral identity for the inverse elements of $N\times N$-matrices
\begin{align}
\label{s1}
    M^{-1}_{nm}=
    \int D(\bar\psi,\psi)\,e^{-\bar \psi M\psi}\psi^\sigma_m\bar \psi^\sigma_n,
\end{align}
where the $2N$ dimensional `graded' vector $\psi=(\psi^\mathrm{b},\psi^\mathrm{f})^T$
contains $N$-component vectors of commuting and Grassmann variables,
$\sigma=\mathrm{b,f}$, respectively. It is set up such that the matrix determinants
generated by the commuting and the Grassmann integral, respectively, cancel out. The
advantage of this design over the alternative replica formalism for the generation of
determinant-free representations is that no spurious analytic continuations are
involved, which provides a more reliable basis for non-perturbative calculations.

Referring to our discussion in the main text, we represent moments of the wave functions as,
\begin{align}
\label{app_moments}
&
I_q
=
\frac{1}{2i\pi \nu_0}\lim_{\delta\to 0}(2i\delta)^{q-1} \sum_n
G^{+(q-1)}_{nn}G^-_{nn},  
\nonumber 
\\
&
G^{\pm}_{nn'}
=
\big\langle n | (\pm i\delta -\hat H)^{-1}|n'\big\rangle, 
\end{align}
where we noted that products of only retarded Green functions do not contribute to the connected average of observables and may be discarded. We now use Eq.~\eqref{s1} with the identification 
 $M=\mathrm{diag}(-i [G^+]^{-1},i [G^-]^{-1})
 =
 -i\sigma_3(
 i\delta
\sigma_3 -\hat H)$ to generate these moments. To this end we introduce 
\begin{align}
{\cal Z}
(\alpha,\beta)
&=
\int D(\bar{\psi},\psi)
e^{-\bar{\psi}
\left(
i\delta \sigma_3 -\hat H
- j(\alpha,\beta)
\right)\psi},
\end{align}
where $\psi=\{\psi^{s,\sigma}_{n}\}$ now is a $4D$ component supervector, carrying
the Fock-space index $n$, and the causal index $s=\pm$ in addition to $\sigma={\rm b, f}$.
The source matrix $j(\alpha,\beta)
=
(\alpha \pi^{\rm b}\otimes\pi^{\rm +} 
+
\beta \pi^{\rm f}\otimes\pi^{\rm -})\otimes |n\rangle\langle n|$
contains  projectors $|n\rangle\langle n|$ in Fock- space and $\pi^{\sigma/s}$ onto subspaces of specific grading/causality, respectively.  
It is now straightforward to verify that the product of 
 resolvents appearing in Eq.~\eqref{app_moments} is represented as
\begin{align}
G_{nn}^{+(q-1)} G^-_{nn}
&={1\over (q-1)!} \partial_\beta\partial_\alpha^{q-1} {\cal Z}
|_{\alpha,\beta=0}.
\end{align}

{\it Effective action:---}Inserting the SYK-Hamiltonian, Eqs.~(1),(2) in the main text, 
the average over couplings $J$ generates a term
quartic in the superfields,
\begin{align}
{\cal Z}(\alpha,\beta)
&=
\int D(\bar{\psi},\psi)
e^{-\bar{\psi}
\hat O \psi - {3J^2\over N^3}\sum_{a}
(\bar{\psi}X_a\psi)
(\bar{\psi}X_a\psi),
}
\end{align}
where $\hat O\equiv i\delta\sigma_3 - \hat H_V - j(\alpha,\beta)$, and $\sum_a$ is a
sum over all ordered index quadruples $a=(i,j,k,l)$, $i<j<k<l$. We reorganize the
quartic term as $(\bar{\psi}X_a\psi) (\bar{\psi}X_a\psi)=-{\rm STr}\left(
\psi\bar{\psi}X_a
\psi\bar{\psi}X_a
\right)$, where the `dyads' $\psi\bar\psi$ represent the composite fields indicated in shading in Fig. 2 of the main text.  
A Hubbard-Stratonovich decoupling in it via a set of  $4D$-dimensional supermatrix fields 
$A_a=\{A^{ss',\sigma\sigma'}_{a,nn'}\}\sim \{\psi^{s\sigma}_n\bar \psi^{s'\sigma'}_{n'}\}$ leads to
\begin{align}
\label{app_HS}
{\cal Z}(\alpha,\beta)
&=
\int DA
e^{-{1\over 2\mathcal{N}}\sum_a {\rm STr}(X_a A_a)^2
+
{\rm STr}\ln\left(
\hat O + i{c_N\gamma \over \mathcal{N}}\sum_a A_a
\right),
}
\end{align}
where  $c_N\equiv 4! {\cal N}/(2N)^4$, $\mathcal{N} \equiv
\left(\begin{smallmatrix}
2N\\4
\end{smallmatrix}\right)$,
 $\gamma=\tfrac{J}{2}(2N)^{1/2}$, and the Gaussian integral
 over $\psi$ has been carried out. 
We
next observe that the `${\rm STr}\ln$' in \eqref{app_HS} couples only to the linear combination
$Y\equiv{1\over {\cal N}}\sum_a A_a$. This motivates a variable change $A_a \mapsto
Y + A_a$, where $\sum_a A_a=0$. Enforcing the constraint via
Lagrange multipliers, it is straightforward to carry out the Gaussian integral over
$A_a$ (see Ref.~\cite{App_ErgodSYK} for a few more details) and to
arrive at 
\begin{align}
\label{app_esm}
{\cal Z}(\alpha,\beta)
&=
\int DY\,
e^{
-{1\over 2}
{\rm STr}(Y{\cal P}^{-1}Y)
+
{\rm STr}\ln\left(
\hat O
+
ic_N\gamma Y
\right)
},
\end{align}
where
${\cal P}Y\equiv {1\over {\cal N}}\sum_a X_a Y X^\dagger_a$.   
 The action of the functional integral defines Eq.~(3) of the main text, 
where we approximated $c_N=1$ and 
the source  has been suppressed, $j(\alpha,\beta)=0$. (The precise value of $c_N=1+\mathcal{O}(N^{-1})$ only enters the comparison with numerics for small system sizes $N=\mathcal{O}(10^1)$.)

 {\it Wave-function statistics:---}Upon projection onto the Fock-space homogeneous mode, 
 Eq.~\eqref{app_esm} reduces to  
the effective action of the Rosenzweig-Porter model (see discussion in main text)
\begin{align}
  \label{app_sm_action}
S[T]
&=
{\rm STr\ln}\left(
i\delta\sigma_3 -\hat H_V - j(\alpha,\beta) + i\gamma \kappa\, Q
\right),
\end{align}
where $Q=T\sigma_3 T^{-1}$. Using the the commutativity of the  fluctuations $T$ with
$\hat H_V$ and cyclic invariance of the `STr-ln' this expression can be
represented as $S[T]
=
{\rm STr\ln}\left(
1
+
G_V{\cal O}_T
\right)$, 
where 
$G^{-1}_V \equiv  i\gamma \kappa\, \sigma_3 - \hat H_V$, 
and 
${\cal O}_T\equiv T^{-1} \left[ i\delta \sigma_3 - j(\alpha,\beta)\right] T$ 
is an operator in which we need to expand to the order required by the correlation function. For example, to linear order in $\delta$ 
one arrives at the action of the zero-dimensional $\sigma$-model 
$S_\delta[Q]
=
\pi\nu_0\delta
{\rm STr}\left(
\sigma_3 Q
\right)$,
where we employed the saddle point solution ${\rm tr}_{\cal F}G_V =
-iY_0=-i\kappa\sigma_3$. In a similar fashion, the source term gives a contribution
$S_j[T]= -
\sum_{k=1}^\infty
g_n^k
\left(
{\alpha^k\over k}
[ Q^{++}_{\rm bb} ]^k
+
\beta 
\alpha^{k-1}
[Q^{++}_{\rm bb}] ^{k-1}Q^{--}_{\rm ff}
\right)$, 
where $g_n\equiv {-i\pi\nu_0\over D(\kappa^2+v_n^2)}$, and we approximated 
$Q^{+-}_{\rm bf}Q^{-+}_{\rm fb}\simeq Q^{++}_{\rm bb}Q^{--}_{\rm ff}$~\cite{appMirlin}. 
Doing the derivatives in the source parameters, we then obtain
\begin{align}
\label{sm_ev}
\partial_\alpha^{q-1} \partial_\beta{\cal Z}|_{\alpha,\beta=0}
&=
g_n^q
q! \,
\langle 
\left[Q^{++}_{\rm bb} \right]^{q-1}
Q^{--}_{\rm ff}
\rangle,
\end{align}

 We finally need to average over the soft mode fluctuation implied by the average $\langle...\rangle=\int dQ \,e^{-S_\delta[Q]}(\dots)$. Referring for details of this  to Ref.~\cite{SMEfetovBook} we here sketch the principal steps of this calculation. The starting point is a `polar coordinate' representation,
 $Q=UQ_0U^{-1}$,
where
$Q_0=\left(\begin{smallmatrix}
\cos\hat{\theta} & i\sin\hat{\theta} \\
-i\sin\hat{\theta}  & -\cos\hat{\theta} 
\end{smallmatrix}\right)$   
is a matrix in causal space 
and we introduced the diagonal supermatrix $\hat{\theta}
={\rm diag}(i\hat{\theta}_{\rm b}, \hat{\theta}_{\rm f})$  
containing 
 compact and  
non-compact angles $0<\theta_{\rm f} <\pi$ 
and
$\theta_{\rm b}>0$, respectively~\cite{SMEfetovBook}.  
 $U$ is a diagonal matrix in causal space which 
contains the four Grassmann variables $\eta^\pm, \bar\eta^\pm$, 
and two more commuting variables  $0\leq\phi,\chi<2\pi$. In this representation, the matrix elements entering the correlation function are given by $Q^{++}_{\rm bb}=\cosh\theta_{\rm bb}(1-4\bar{\eta}^+\eta^+)$ and
$Q^{--}_{\rm ff}=\cos\theta_{\rm ff}(1-4\bar{\eta}^-\eta^-)$, and  
the integration 
measure reads
$dQ={1\over 2^6\pi^2}{\sinh\theta_{\rm b}\sin\theta_{\rm f}\over (\cosh\theta_{\rm b}-\cos\theta_{\rm f})^2}
d\phi d\chi d\theta_{\rm b}d\theta_{\rm f} d\bar{\eta}^+d\eta^+d\bar{\eta}^-d\eta^-$~\cite{SMEfetovBook}.    
The essential advantage of the polar representation is that the  action only depends on the `radial variables' 
$S_\delta[Q]=2\pi\nu_0\delta(\cosh\theta_{\rm b}-\cos\theta_{\rm f})$. This shows how the integration over the non-compact angle is cut by the parameter $\delta$ at  values 
$1\leq \lambda \equiv \cosh\theta_{\rm b}\lesssim 1/\delta$, while 
the integration over the compact angles $\theta_\mathrm{f}$ is free. With these structures in place, it is straightforward to obtain 
\begin{align}
G_{nn}^{+(q-1)} G^-_{nn}
&\simeq
2q(q-1)g_n^q
\int_0^\infty d\lambda\, 
\lambda^{q-2} e^{-2\pi\nu_0\delta\lambda}. 
\end{align}
Doing the final integral and collecting all factors
we arrive at Eq.~(6) in the main text.

{\it Level-level correlations:---} The two level correlation function $K(\omega)=\nu_0^{-2} \langle \nu(\omega/2)\nu(-\omega/2)\rangle$ probing spectral statistics in the middle of the band can be obtained in similar ways. Starting from the representation, 
\begin{align}
K(\omega)
&=
{1\over \pi^2\nu_0^2}
\langle {\rm Im}\, {\rm tr}\, G(\tfrac{\omega}{2}) {\rm Im}\, {\rm tr}\,G(-\tfrac{\omega}{2})\rangle,
\end{align}
we introduce a  source matrix 
 $j(\alpha,\beta)
=
(\alpha \pi^{\rm b}\otimes\pi^{\rm +} 
+
\beta \pi^{\rm f}\otimes\pi^{\rm -})$ into the action generalized for finite
frequency differences, $i\delta \sigma_3 \to (\omega/2+i\delta)\sigma_3$ in Eq.(3) of
the main text. From this representation, the function $K$  is obtained  as
$K(\omega)={1\over 2\pi^2\nu_0^2} {\rm Re}\,
\partial^2_{\alpha\beta}{\cal Z}|_{\alpha,\beta=0}$. 
Proceeding as in the computation of the wave-function moments,  
one obtains~\cite{SMEfetovBook}
\begin{align}
K(\omega)
&=
{1\over 2}{\rm Re}
\int_1^\infty d\lambda_{\rm b}
\int_{-1}^1 d\lambda_{\rm f}\,
 e^{i\pi\nu_0(\lambda_{\rm b}-\lambda_{\rm f})},
\end{align}
where $\lambda_{\rm b}\equiv \lambda$ is the non-compact bosonic angle introduced above and 
 $\lambda_{\rm f}$ the compact fermionic angle. These integrals can be carried out in closed form, and yield the two-point correlation function of the Gaussian Unitary Ensemble,
 $K(s)=1-\tfrac{\sin^2(\pi s)}{(\pi s)^2}+\delta(s)$, with $s=\omega\nu_0$. This demonstrates that the presence of an inhomogeneous diagonal in the Green functions affects the wave function moments, but not the spectral correlation functions~\cite{SMOssipov}. The reason for this is that the wave function moments --- formally obtained as powers of Green function matrix elements --- respond more sensitively to the presence of a fluctuating diagonal than the spectral correlation function --- obtained via tracing over single Green functions.



\begin{thebibliography}{99}
\bibitem{fn1}
We here consider values $2N=2,6 \,\mathrm{mod}\,8$ where the SYK model is in the unitary
symmetry class~\cite{GG2}. 
The description of other cases requires $8$-dimensional
integration matrices reflecting the symmetry of the orthogonal or symplectic symmetry
class, respectively.
\bibitem{fn2}
The precise definition is
$F(x)=\tfrac{2}{\pi(1+x^2)}+\tfrac{2}{\pi x}\arctan(x)-\tfrac{4}{\pi
x^2}\arctan^2(x)$, and we used that $\sum_m \Pi^{-2}_{nm}\simeq 42/N^4$.
\bibitem{fn3}
With $N=13$,  $\kappa\simeq \pi/\Delta$, and the identification $\kappa^2={20\pi\over N^4\kappa\Delta_P}$ stated above we obtain  $\Delta_{\rm P}\simeq {169 \pi \over \sqrt{20}}\simeq 119$.
\end{thebibliography}
\end{document}